\documentclass[12pt]{article}

\usepackage{amsfonts,amssymb}
\usepackage{amsmath}

\usepackage{float}
\usepackage{enumerate}
\usepackage{supertabular,booktabs}
\usepackage{threeparttable}
\usepackage{multirow}
\usepackage{lscape}
\usepackage{array}
\usepackage[ruled]{algorithm2e}

\title{On the Relational Translation Method for Propositional Modal Logics\thanks{Supported in part by the National High Technology Program under grant No. 863-306-05-03-3.}}

\author{
Jian Zhang \\
Laboratory for Computer Science \\
Institute of Software \\
Chinese Academy of Sciences \\
P.O.Box 8718, Beijing 100190, P.R. China 
}

\date{December 2021\thanks{Reprinted (with a few minor changes)
with permission from Technical Report ISCAS-LCS-96-12, Laboratory of Computer Science, Institute of Software, Chinese Academy of Sciences, Dec. 1996.
Some of the results were published as a short paper in Chinese:
``A Translation Approach to Modal Logic Reasoning'',
{\it Computer Research \& Development}, Vol. 35, No. 5, pp.389--392, 1998.}}

\begin{document}




\setcounter{page}{1}

\maketitle

\begin{abstract}
One way of proving theorems in modal logics is translating them into the predicate calculus and then using conventional resolution-style theorem provers. This approach has been regarded as inappropriate in practice, because the resulting formulas are too lengthy and it is impossible to show the non-theoremhood of modal formulas. In this paper, we demonstrate the practical feasibility of the (relational) translation method. Using a state-of-the-art theorem prover for first-order predicate logic, we proved many benchmark theorems available from the modal logic literature. We show the invalidity of propositional modal and temporal logic formulas, using model generators or satisfiability testers for the classical logic. Many satisfiable formulas are found to have very small models. Finally, several different approaches are compared.
\end{abstract}

\vspace{5mm}

\section{Introduction}

Modal logics have been studied by many researchers in computer science and artificial intelligence. There are roughly two classes of methods for automated reasoning in these logics. With the {\em translational\/} approach, we transform modal formulas to first-order predicate logic (FOPL) formulas and then use existing tools for the classical logic. Alternatively, we can design tools specifically for modal logics. Such methods are classified as the {\em direct\/} approach. They include, among others, semantic tableaux, modal resolution, and matrix-based procedures. See, for example, \cite{ref1, ref5, ref22}.

The translational approach can be further divided into syntactic and semantic methods \cite{ref16}. To use the {\it syntactic\/} method, we introduce a special unary predicate $P$ which means its argument is provable, and translate the axioms and rules into classical logic. This method can only be applied to propositional modal logics. On the other hand, the {\it semantic\/} methods are based on Kripke semantics. Such a method is applicable to first-order modal logics, but properties of the accessibility relation should be defined by a finite set of sentences in predicate logic. This can be done in a number of ways. In the standard way \cite{ref16}, we introduce a special binary predicate $R$ to describe the accessibility relation between the possible worlds.
This is now called {\it relational\/} translation. In addition, there are so-called {\it functional\/} translation and {\it semi-functional \/} translation \cite{ref18, ref17}.

The relational translation approach is very simple and quite general. It can deal with many modal logics, and it benefits from the power of existing tools for the classical logic. But there are also some problems with it. To quote Gor\'{e} \cite{ref8} (page 2), ``by translating into first order logic, the translational methods immediately surrender the decidability of the propositional modal logic they translate.''
Another problem is that, the result of translation is usually a complicated formula, which makes its proof difficult to find. Ohlbach and Weidenbach \cite{ref19} gave the following example:
\begin{center}
Prove that $(\Diamond \Box p \leftrightarrow \Diamond \Box \Diamond \Box p)$ is valid in {\bf KD45}.
\end{center}
\noindent
Using the standard translation method and the popular resolution-style prover OTTER 3.0 \cite{ref13}, they failed in finding a proof.

For these reasons, the (relational) translation method has been neglected by many researchers. Is it really that bad? Our experiments show that it is quite competitive. In this paper, we shall describe some of the results. We suggest using both theorem provers and counterexample finders to decide the validity of modal formulas.
In addition, we study the problem of satisfying propositional temporal logic formulas.

\section{Translating Normal Modal Logics}

\subsection{Propositional Normal Modal Logics}

We first review some relevant definitions and facts. For more details, see \cite{ref2, ref10}. There are many modal logic systems. But in this paper, we shall consider only normal modal logics.

The smallest normal system is called {\bf K}. It can be extended by defining additional axioms. The most well-known axiom schemas include {\it D, T, B, 4, 5}.
There are only 15 distinct normal systems (including {\bf K} itself) produced by taking the schemas (as axioms) in all combinations. They are named {\bf K, KD, KT4,} and so on. Among them, {\bf KT4} and {\bf KT5} are also known as {\bf S4} and {\bf S5}, respectively.

The semantics of modal logics is often defined in terms of the possible worlds structures. Many propositional modal systems have the {\em finite model property\/}, which means, every non-theorem is false in some finite model. This property implies decidability, if the logic is finite axiomatizable. It is well known that each of the above 15 modal logics has the finite model property. Moreover, in some systems (e.g. {\bf S5}), any satisfiable formula can be satisfied in a small finite model (whose size does not exceed the length of the formula).

\subsection{The Relational Translation Method}

Roughly speaking, the modal operators $\Box$ and $\Diamond$ correspond to universal and existential quantifiers in FOPL respectively. There is a standard way of translating modal formulas into FOPL formulas. 
To do so, we add a `world' argument to each proposition, and introduce a new binary predicate $R$ to describe the accessibility relation between worlds. If the current world is $w$, then the formula $\varphi$ will be translated to $Tr(\varphi,w)$, which means $\varphi$ is true at $w$. The basic translation rules are defined inductively as follows:
\begin{align*}
    Tr(p,w) & ~\equiv~ p(w)\\
    Tr(\neg \psi, w) & ~\equiv~ \neg Tr(\psi,w)\\
    Tr(\psi_1 \wedge \psi_2, w) & ~\equiv~ Tr(\psi_1,w) \wedge Tr(\psi_2,w)\\
    Tr(\Box \psi, w) & ~\equiv~ \forall v (R(w,v) \rightarrow Tr(\psi,v))\\
    Tr(\Diamond \psi,w) & ~\equiv~ \exists v (R(w,v) \wedge Tr(\psi,v))
\end{align*}
where $v$ is a new `world' variable. The proposition $p$ in the original formula becomes a unary predicate.
For example, $Tr(\Diamond \Box p, o)$ is the formula
$$ \exists w (R(o,w) \wedge \forall v (R(w,v)\rightarrow p(v))) $$

The following theorem \cite{ref16,ref15} forms the basis of the relational translation method.

\textbf{Theorem.} A formula $\varphi$ is valid in a modal logic system $S$ if and only if $Ax(S) \rightarrow \forall w Tr(\varphi,w)$ is valid in first-order predicate logic, where $Ax(S)$ is a set of axioms describing properties of the accessibility relation.

For example, $(\Box p \rightarrow \Diamond p)$ is valid in {\bf KT}, because the sentence 
\begin{align*}
    \forall x & R(x,x) ~\rightarrow \\
    \forall w & (\forall v (R(w,v) \rightarrow p(v)) ~\rightarrow~ \exists v (R(w,v) \wedge p(v)))
\end{align*}
is a theorem in the predicate calculus.

\subsection{A Concurrent Program as the Decision Procedure}

The concurrent program in Algorithm 1 consists of a theorem proving process ($P_1$) and a model finding process ($P_2$).
The theorem proving process can be based on any refutationally complete proof procedure. To find a model of some fixed size, one can either use a decision procedure for the classical propositional calculus (such as the Davis-Putnam algorithm) or use a finite model search program for the first-order predicate logic.
The input parameter $\varphi$ is a propositional modal logic formula and $S$ is a normal modal logic system.

We assert that the concurrent program can serve as a decision procedure for many propositional modal logics like \textbf{K}, \textbf{KT}, \textbf{KD}, \textbf{S4}, \ldots. Such a modal system should possess the finite model property, and the accessibility relation should be characterized by a finite set of sentences in first-order predicate logic. The program terminates under the fairness assumption. (It is unfair if one process, e.g., the theorem prover, never has a chance to be executed.)

\begin{algorithm}
\caption{The decision procedure CDP}
	\KwIn{$\varphi$, $S$} 
	\textbf{cobegin} \\
	$\phi:=Ax(S) \rightarrow \forall w \ Tr(\varphi, w)$ \;
	$P_1 ::$ \\
	\Repeat{(contradiction is deduced)}{
	apply a suitable set of inference rules to $\phi$\;
	}
	\textbf{kill}$(P_2)$\;
	$P_2::$ \\
	\textbf{var} $n=0$\;
	\Repeat{(a model is found)}{
	 $n:=n+1$\;
	 find an $n$-element model of $\phi$\; 
	}
	\textbf{kill}$(P_1)$\;
	\textbf{coend}
\end{algorithm}

\section{Experimental Results}

For non-classical logics, there are not so many automated reasoning tools and test problems. Here we describe some experimental results on the benchmarks used by other authors \cite{ref1,ref3,ref9,ref6}.
We shall not elaborate on the programs' performances, because they are affected by several factors such as the data structures and the programming languages. Moreover, not all timing information are available in the related papers.

It is very easy to implement a tool for translating modal logic formulas. (The translation time will be neglected.) To show the satisfiability of the formulas, we may use various tools, such as FINDER, MGTP, LDPP, SATO, MACE and SEM \cite{ref20,ref21,ref23,ref14,ref25}.
In the following, we only report the size of the smallest model satisfying each formula. In many cases, the models are very small, and can be easily found by any of the tools.

To prove modal theorems, we use the resolution-style theorem prover OTTER 3.0.4 \cite{ref13}, running on a SPARCstation 20 with 32 MB memory. We did not take much advantage of OTTER's special features to achieve high performances. It was instructed to run in autonomous mode. However, since the prover is not so good at proving
``if-and-only-if'' theorems \cite{ref12}, we broke each theorem of the form $A \leftrightarrow B$ into two cases $A \rightarrow B$ and $B \rightarrow A$.
In such a way, Ohlbach and Weidenbach's example mentioned earlier can be proved within 2 seconds.

\subsection*{Problem Set 1\footnote{Some early results on this set of problems were described in \cite{ref24}.}}

In \cite{ref1} Catach describes a tableaux-based program called TABLEAUX, and gives the validity status of 31 formulas in 16 modal logic systems. The formulas are very simple, and TABLEAUX completed all 496 tests in less than 1 minute. For simplicity, we consider only the logics \textbf{K}, \textbf{KD45}, \textbf{S4} and \textbf{S5}, which are very important in knowledge representation and reasoning.

There are totally $31 \times 4 = 124$ tests to be performed. Among these, 67 cases are validity proofs. With only a few exceptions, OTTER finds each proof within 1 second. In the remaining 57 cases, we found the smallest structures which show the invalidity of the formulas. Of these 57 countermodels, only 8 are of size 3. The others are of size 1 or 2.

For those formulas containing the equivalence connective, if we give them directly to the translator and then to OTTER, it will be much more difficult to find the proofs. For example, the last formula
$\Box \Box  p \leftrightarrow \Diamond \Box p$ is valid in \textbf{KD45}.
It takes OTTER about 2 minutes to find a proof. But each of the formulas $\Box \Box p \rightarrow \Diamond \Box p$ and $\Diamond \Box p \rightarrow \Box \Box p$ can be proved to be valid in less than 0.25 second.

\subsection*{Problem Set 2}

Demri \cite{ref3} analyses several different strategies in a tableau-based \textbf{S4} prover, and compares them on a set of 9 valid formulas. Here we list 4 of them:
\begin{itemize}
    \item[4.]$\Box_{c} (\neg PC \rightarrow \Box_{b} \neg PC)$\\
            $\wedge \Box_c \Box_b \Box_a (PC \vee PB \vee PA)$ \\
            $\wedge \Box_c \Box_b (\neg PB \rightarrow \Box_a \neg PB)$ \\
            $\wedge \Box_c \Box_b (\neg PC \rightarrow \Box_a \neg PC)$ \\
            $\wedge \Box_c \neg \Box_b PB$ \\
            $\wedge \Box_c \Box_b \neg \Box_a PA$ \\
            $\rightarrow \Box_c PC$ 
    \item[5.] $\Diamond \Box (\Box(p \vee \Box q) \leftrightarrow (\Box p \vee \Box q))$
    \item[6.] $\Diamond \Box ((p \rightarrow q) \leftrightarrow F(q,F(p,q)))$\\
    where $F(A,B)$ stands for $(\neg A \vee \neg \Diamond(A \wedge B) \vee (B \wedge \Diamond(A \wedge \neg B)))$
    \item[9.] $\Box(\Box(\Box p \rightarrow \Box(\Box q \rightarrow \Box r)) ~\rightarrow~ \Box(\Box(\Box p \rightarrow \Box q)\rightarrow \Box r))$
\end{itemize}
The first one is a multimodal formula, encoding McCarthy's 3 Wise Men puzzle. (C is the wisest man.) There are typos in formulas (6) and (9). The correct versions are as follows \cite{ref4}:
\begin{itemize}
\item[6a.] $\Diamond \Box ((p \rightarrow q) \leftrightarrow F(p,F(p,q)))$
\item[9a.] $\Box(\Box(\Box p \rightarrow \Box(\Box q \rightarrow \Box r)) ~\rightarrow~ \Box(\Box(\Box p \wedge \Box q)\rightarrow \Box r))$
\end{itemize}
Formulas (5) and (6a) are obtained from \textbf{S5}-valid formulas. A formula $\phi$ is satisfiable in \textbf{S5} iff $\Diamond \Box \phi$ is satisfiable in \textbf{S4}. 
Instead of proving the validity of $\Diamond \Box \phi$ in \textbf{S4}, we prove directly $\phi$ is valid in \textbf{S5}.
As previously, we divide each ``if-and-only-if'' theorem into two cases. In this way, 8 of the 9 theorems are easily proved, each requiring less than 2 seconds. The exception is formula (6a) whose proof is found in about 1 minute.

It is interesting to note that some non-theorems are falsified in very small models. For example, the negation of (6) and (9) are satisfied in 1-world models. And, if we substitute $\Box_b PB$ for $\Box_c PC$ in formula (4), then there is also a countermodel of size 1, which can be easily found by exhaustive search.

\subsection*{Problem Set 3}

Recently Gor\'{e}, Heinle and Heuerding \cite{ref9} studied the relations between some propositional normal modal logics, using the Logics Work Bench (LWB). As a side-effect of this work, they collected a database of theorems which can be used to test modal theorem provers.

LWB has automated proof procedures for only a few modal logics, namely, \textbf{K}, \textbf{KT}, \textbf{KT4}, \textbf{KT45} and \textbf{KW}. To deal with extensions of these logics, one can add some modalised instances of the new axioms to the premises. The user has to provide the appropriate instances. For more details about this technique, see \cite{ref9}.

\vspace*{3mm}

In the appendix of \cite{ref9}, 72 K-theorems are listed. Most of them are easy for OTTER to prove, but there are also some difficult ones. The results are summarized in the following table.
\begin{table}[h]
    \centering
    \begin{tabular}{c|c}
    \hline
    OTTER time & number of theorems \\
    \hline
    \textless 10 sec.  &  51  \\
    10-100 sec.  &  4   \\
    100-1000 sec.  &  5  \\
    \textgreater 1000 sec.  &  12  \\
    \hline
    \end{tabular}
\end{table}

So about 3 quarters of the theorems can be proved within 2 minutes. The 12 difficult formulas are very complicated. Some of them can be easily proved if we work in a different modal system rather than in \textbf{K}. For example, using our translation tool and OTTER, we can prove that $M \rightarrow Pt$ is valid in \textbf{K4} within 10 seconds.
Here $M$ and $Pt$ stand for the following two formulas:

$$  M ~\equiv~ \Box \Diamond p \rightarrow \Diamond \Box p \ \ \ \ \ Pt ~\equiv~ \Box(p \vee \Diamond p) \rightarrow \Diamond(p \wedge \Box p) $$

Instead of proving the theorem directly, Gor\'{e}, Heinle and Heuerding prove the following \textbf{K}-theorem:
$$\phi_1 \wedge \phi_2 \wedge \phi_3 \wedge M \rightarrow Pt$$
where $\phi_1$ is an instance of \emph{4}, $\phi_2$ and $\phi_3$ are instances of $\Box$\emph{4}. 
The resulting formula is quite lengthy, and OTTER has difficulty finding a proof for it. I am not indicating that the ``modalised instance'' technique is not useful. But the generality and power of the translation approach should be emphasized. Of course, a powerful prover for the predicate calculus is very important.

Based on the formulas given in \cite{ref9}, we also examined some conjectures which are not valid in \textbf{K}. They are usually falsified in very small Kripke structures. The following are some examples:
\begin{table}[h]
    \centering
    \begin{tabular}{c|c}
    \hline
    formula & size of structure \\
    \hline
    $ M \rightarrow Pt $  &  2 \\
    $ H \rightarrow L $  &  3 \\
    $ H^+ \rightarrow L^+ $  &  2 \\
    $ L \rightarrow L^+ $  &  2 \\
    $ L^{++} \rightarrow L^+ $  &  2 \\
    $ Dum4 \rightarrow Dum $  &  2 \\
    \hline
    \end{tabular}
\end{table}

The formulas $H$, $H^+$, $L$, $L^+$ and $L^{++}$ are defined as follows:
\begin{align*}
    H & ~\equiv~ (\Box(p \vee q) \wedge \Box (\Box p \vee q) \wedge \Box(p \vee \Box q)) \rightarrow (\Box p \vee \Box q)\\
    H^+ & ~\equiv~ (\Box(\Box p \vee q) \wedge \Box(p \vee \Box q)) \rightarrow (\Box p \vee \Box q)\\
    L & ~\equiv~ \Box((p \wedge \Box p) \rightarrow q) \vee \Box((q \wedge \Box q)\rightarrow p)\\
    L^+ & ~\equiv~ \Box(\Box p \rightarrow q) \vee \Box(\Box q \rightarrow p)\\
    L^{++} & ~\equiv~ \Box(\Box p \rightarrow \Box q) \vee \Box(\Box q \rightarrow \Box p) \\
    Dum & ~\equiv~ \Box(\Box(p \rightarrow \Box p) \rightarrow p) \rightarrow (\Diamond \Box p \rightarrow p) \\
    Dum4 & ~\equiv~ \Box(\Box(p \rightarrow \Box p ) \rightarrow p) \rightarrow (\Diamond \Box p \rightarrow (p \vee \Box p))
\end{align*}

To show that $Dum$ holds in \textbf{KTDum2}, Gor\'{e}, Heinle and Heuerding proved the \textbf{K}-theorem:
$$ \varphi_1 \wedge \varphi_2 \wedge \Box Dum2 \rightarrow Dum $$
where $\varphi_1$ and $\varphi_2$ are instances of $T$ and $\Box T$, respectively. $Dum2$ is defined by :
$$ Dum2 ~\equiv~ \Box(\Box(p \rightarrow \Box p ) \rightarrow \Box p) \rightarrow (\Diamond \Box p \rightarrow p)$$
We tried to find a small reflexive structure falsifying $Dum2 \rightarrow Dum$, but failed.
The formula turned out to be \textbf{KT}-valid, which can be proved in about 1 second.
So $Dum$ is implied by either $Dum2$ or $\Box Dum2$ in \textbf{KT}.

\subsection*{Problem Set 4}

Giunchiglia and Sebastiani \cite{ref6} developed a SAT-based decision procedure for \textbf{K}, called KSAT. It has been tested against a large number of randomly generated 3CNF modal formulas. We implemented a similar method for generating random formulas, and found that quite some satisfiable formulas have very small models. But the number of formulas that have been tested is not enough for us to make any conclusion.

We did not spend much time on this set of problems. There are two reasons for this. Firstly, the generated formulas usually contain repetitive or complementary literals in the same clause, e.g. $\Box(p_1 \vee p_2 \vee \neg p_1)$.
This problem is quite serious when there are only a few variables (say, fewer than 5). Secondly, we are more interested in structured formulas. But when only the average timings are recorded, some hard formulas
(e.g., formulas of the form $ A \leftrightarrow B $) can be hided in many other easy formulas.

\section{Satisfying Propositional Temporal Logic Formulas}

Temporal logic was introduced into computer science by Amir Pnueli about 20 years ago. It is a convenient tool for specifying and verifying concurrent programs \cite{ref11}. There are many versions of temporal logic. For example, time may be discrete or dense, a time point can have only one successor or several successors. Here we consider the linear time temporal logic (LTL), where time is modeled by the set of natural numbers.
In addition to the `box' and `diamond' operators, LTL has such operators as `next-time' and `until'.

Since LTL is based on the natural numbers, induction is needed to prove some theorems. A resolution-style automatic theorem prover is not enough if we translate LTL formulas into FOPL formulas. However, propositional LTL also has the finite model property, i.e., a formula is satisfiable iff it has a model consisting of a finite number of states. And we can still use model generation tools in FOPL to find models of satisfiable propositional temporal logic formulas. But we have to check the generated model can be translated into some LTL model.

For simplicity, let us consider only the next-time operator. In this case, we need a new unary successor operator $S$. The binary relation $R$ is reflexive and transitive. In addition, we have to add such axioms as $R(x,S(x))$.

Let us give a simple example to illustrate the difference between models found in this way and models obtained by the tableau method. Consider the formula $\bigcirc \bigcirc \Box p$, where $\bigcirc$ denotes next-time. With the tableau method, we get a 3-state model:
$\{s_0, s_1, s_2\}$, where $S(s_0)=s_1$, $S(s_1)=s_2$, $S(s_2)=s_2$, $p$ can take any value in the first two states, but it is true in $s_2$.
In general, if there are $k$ next-time operators, we shall get a $(k+1)$-state model.
However, with the translation method, a 1-state model is produced:
$\{s_0\}$, $S(s_0)=s_0$, $p$ holds at $s_0$.

\section{Related Work}

As we mentioned in the introduction, there are quite some methods for the automated reasoning in modal logics. But relatively few implementations are available. The matrix method and the functional translation method require specialized unification algorithms. In contrast, it is very easy to write a relational translator.

Another advantage of the translational approach is its generality. It can be applied to many modal logic systems. Changing from one system to another, we need only modify the axioms. On the other hand, many other methods and tools are designed for a few specific systems. For example, semi-functional translation \cite{ref17} and the technique described in \cite{ref19} are most suitable for {\it serial\/} modal logics like \textbf{KD} and \textbf{KD45}.
Modal resolution rules are discussed in \cite{ref5}, within 5 systems, i.e., \textbf{K, Q, T, S4} and \textbf{S5}.
The program LWB has built-in proof procedures for 5 selected logics, and KSAT \cite{ref6,ref7} works only in \textbf{K}.
TABLEAUX can deal with many modal systems. But if one was to extend it with new systems, new procedures (encoding the semantic properties) would have to be added.

Tableau-based decision procedures can be used both to prove theorems and to show the satisfiability of formulas. As we point out in the previous section, the models found by our method are usually different from those generated by tableau procedures. An example in linear time temporal logic has been given to show the difference.
For another example, the formula \emph{4}. $\Box p \rightarrow \Box \Box p$ is not valid in \textbf{K}.
Catach \cite{ref1} gives a 3-world countermodel produced by the tableau-based procedure. But with our method, we shall first find the smallest model having 2 worlds:

\begin{table}[H]
\centering
\begin{tabular}{p{0.5cm}<{\centering}|p{1cm}<{\centering}p{1cm}<{\centering}p{0.5cm}<{\centering}p{0.5cm}<{\centering}p{1cm}<{\centering}p{1cm}<{\centering}}
$R$ & $w_0$ & $w_1$ &  & \multicolumn{1}{c|}{$p$} & $w_0$ & $w_1$ \\ \cline{1-3} \cline{5-7} 
$w_0$ & $False$ & $True$ &   & \multicolumn{1}{c|}{}  & $False$ & $True$ \\
$w_1$ & $True$ & $False$ &   &    &    &   
\end{tabular}
\end{table}
\noindent where $w_0$ is the real world.

According to our experiences, many satisfiable formulas have very small models. Thus it is practical to use model generation tools in the classical logic to show the satisfiabiliy of propositional modal formulas. However, for some modal logics (like \textbf{K}), there are some specially constructed formulas \cite{ref10} which are satisfiable only in exponential-size models. They present much difficulty for methods based on the Kripke semantics. The program KSAT can deal with such formulas efficiently \cite{ref7}.

\section{Concluding Remarks}

The (relational) translation technique has been known for a long time. But few reports on its practical performances are available, and previously no one has studied the generation of countermodels with this method. Our experiments show that, the method is quite competitive. Without spending much effort, we can prove a lot of theorems in many modal logics, using existing tools in the classical logic. Moreover, by combining a theorem prover with a model generator, we can decide the satisfiability as well as the validity of propositional formulas. It is interesting that many non-theorems turn out to have very small countermodels.

Compared with other methods, relational translation has some appealing features. (See the previous section.) Theoretically, it provides decision procedures for various important logics. Practically, we expect that it can deal with a large number of formulas of reasonable sizes. It should be very useful in those applications where first-order reasoning is inevitable, and in the applications which have several modal operators belonging to different systems.

It is certainly not true that the translational approach is better than other approaches in all respects. The method has some drawbacks. But it is not so weak as thought previously. Rather than proposing an entirely new method, we try to identify and overcome some difficulties. The power of the translation approach is largely dependent on the tools available in predicate calculus. In our experiments, only OTTER (in its autonomous mode) was used to prove theorems. This may not be the best choice. One can use other types of provers (e.g., a nonclausal one) or exploit more features of OTTER to achieve better performances. One can also develop special techniques for handling the transitivity axiom or the seriality axiom (similar to that of \cite{ref19}).

\end{document}